# Ranking Search Engine Result Pages based on Trustworthiness of Websites

Srikantaiah K C[1], Srikanth P L[1], Tejaswi V[1], Shaila K[1], Venugopal K R[1] and L M Patnaik[2]

[1] Department of Computer Science and Engineering
University Visvesvaraya College of Engineering,
Bangalore University, Bangalore-560 001, India

[2] Honorary Professor, Indian Institute of Science,
Bangalore, India

**Abstract**
The World Wide Web (WWW) is the repository of large number of web pages which can be accessed *via* Internet by multiple users at the same time and therefore it is Ubiquitous in nature. The search engine is a key application used to search the web pages from this huge repository, which uses the link analysis for ranking the web pages without considering the facts provided by them. A new algorithm called *Probability of Correctness of Facts(PCF)-Engine* is proposed to find the accuracy of the facts provided by the web pages. It uses the Probability based similarity function (SIM) which performs the string matching between the true facts and the facts of web pages to find their probability of correctness. The existing semantic search engines, may give the relevant result to the user query but may not be 100% accurate. Our algorithm computes trustworthiness of websites to rank the web pages. Simulation results show that our approach is efficient when compared with existing Voting and Truthfinder[1] algorithms with respect to the trustworthiness of the websites.

## 1. Introduction

WorldWide Web (WWW) is a collection of interconnected web pages accessed *via* internet which offers information and data from all over the world. When searching for a topic in the WWW, it returns many links or web sites related on the browser to a given topic. The important issue is to determine the website that gives the accurate information. There are many related web sites that give unauthoritative information. While the information in other repositories like books, library and journals is evaluated by scholars, publishers and subject experts. We have no mechanism to evaluate the information on WWW. Hence, it is necessary to consider some criteria[2] to evaluate the information hosted on WWW.

Web Search Engines are programs used to search information on the WWW and FTP servers and to check the accuracy of the data automatically. Web search engines are classified into following categories: *Crawler based*, Directories, *Hybrid Engines*, *Meta Engines* and *Specialty Search Engines*. Crawler based search engines use crawler to survey and categorize the web pages (example google.com), Directories use manual editors to survey and categorize the web pages. Human editors map web sites to specific categories in the directory database, based on the information they find and uses a set of predefined rules, for example: yahoo directory(www.yahoo.com) and open directory (www.dmoz.org).

Hybrid Search Engine makes use of properties of crawler based search engine and directories i.e., use a combination of crawler based results and directory results, example modern search engines like yahoo (www.yahoo.com) and google (www.google.com). Meta Search Engine combines the results from all other search engine into one large listing, example meta crawler (www.metacrawler.com) and specialty search engines are used to search in specific area, such as shopping(www.shopping.yahoo.com). It operates in the following order: Web Crawling, Indexing, Searching and Ranking.

Web crawling is a process of extracting new pages by a program called spider or robot. Indexing is a process of extracting an important words from the title, heading and Meta tags of a newly crawled page. These words are stored in the Index Database, will be used by later queries. Searching is a process of searching web pages in the index database corresponding to the keywords, entered by the user in the search engine. The search engine examines its index database and returns a listing of best matching web pages (result set) based on its criteria.

Ranking is a process of arranging the retrieved WebPages of the Search Engine Result Page (SERP) based on the relevance of the query entered. Relevance of a webpage is calculated based on the contents of the web page, including title, Meta data, popularity, authority, facts, location and frequency of a term in a web page.

*Motivation:* The existing page rank algorithms such as Authoritative – Hub analysis and PageRank uses the statistical analysis i.e., the rank for the page is calculated based on the number of links referring to the page and on the importance of the referring pages. The facts provided by the web pages are not considered while assigning the ranks to the pages.

*Contribution:* A new approach called PCF-Engine is proposed in this paper to find the trustworthiness of websites using probability of correctness of facts which is found by applying Probability based similarity function (*SIM*) between conflicting facts and true facts available in the knowledge base.

*Organization:* The rest of this paper is organized as follows: Section 2 and 3 describe Related Work and Background respectively. Section 4 defines problem, describes Mathematical Model and algorithm, section 5 describes efficiency of PCF-Engine algorithm. Section 6 describes the system architecture. Section 7 comprises of experimental results and analysis. The concluding remarks are summarized in Section 8.

## 2. Related Work

Several algorithms have been proposed to rank the WebPages retrieved by search engine and they are categorized into Authority Based and Fact Based. Authority based algorithms model the entire web as Web Graph, where nodes represent the WebPages and edge represents the hyperlink between the two pages, the node with maximum indegree is considered as most authoritative and is ranked at the first position. But most authority pages do not contain accurate information [3].

PageRank[4] and HITS [5] are authority based rank algorithms. PageRank Algorithm has been developed by Larry Page, ranks the webpages based on the indegree of a node in the web graph and it is a query independent algorithm. Further, the original pageRank algorithm has been improved by considering weights to links, Cluster Prediction, Subgraphs, Timestamp, Extrapolation method, Index, and Machine Learning.

Wenpu Xing et al., [6] propose weighted pagerank algorithm, which is the extension of standard pagerank algorithm. It calculates the popularity of the webpages using both inlinks and outlinks of the pages in the reference page list, but multiple levels of reference pagelist are not considered. It outperforms theconventional algorithm in terms of returning large number of relevant pages to a given query.

DRANK[7] models web graph as a three layer graph which includes host graph, Directory graph and page graph by using hierarchical structure of URLs and structure of link relation of the web pages. It predicts future ranking values of pages using clustering technique. Yenyu Chen et al.,[8] address local methods for estimating pagerank values. It computes pagerank score of particular webpage using only small subgraph of the entire graph. Heason Hwang et al.,[9] have proposed a binrank algorithm, it ranks the high quality search results. High quality search results are retrieved by generating subgraphs by partitioning all the terms as corpus based on their co-occurrence.

Shigang Ju et al., [10] introduce WTPR algorithm, which is based on temporal link analysis and it is fit for dynamic changes of the net. WTPR can make old pages decline and new pages raise in the resulting page, meanwhile it can help the old pages which have high quality get higher rank than common old pages. Sepandar D Kamvar et al.,[11] propose a method for reducing time required to calculate the rank scores of power method by 25-300%. It accelerates the process by periodically subtracting off estimates of the non principal eigen vectors from the current iterate of the power method.

Yong Zhang et al.,[16] propose an algorithm A-pagerank, in which the pagerank of the source page is distributed to its linkout pages according to the topic similarity. Amit Pathak et al.,[12] present an algorithm to reduce time and increase the ranking of the popular pagerank algorithm, by using indexing technique. Sweah Liang Yong et al.,[13] propose a generic webpage ranking framework based on Graph Neural Networks (GNN) and on all existing pagerank algorithms. GNN can be used as a more ranking mechanism to the existing numerical ranking method for webpages. The Hypertext Induced Topic Search (HITS) and its variants [14,15], rank the webpage based on both indegree and outdegree of a node in the web graph and it is a query dependent algorithm.

Matthew Bennett et al.,[14] propose an efficient and scalable implementations of the HITS algorithm called Parallel HITS(PHITS), it uses shared memory supercomputer.

Brian amento et al.,[15] evaluate the performance of many ranking algorithms, which are based on link analysis, using dataset of less domain and topic, web documents rated for quality by human topic experts. Link based metrics did a good job by improving precision about 0.55 to 0.75. Bo Zhang et al.,[17] develop a Semantic based Trust Computation Scheme, in which trustable semantics information is represented through ontology. Entire trust is measured by a combined trust score from both subjective and objective sides of information.

Truthfinder[1] is a fact based search engine; it ranks websites by computing trustworthiness score of each website using the confidence of facts provided by websites. It utilizes the relationships between websites and their information to find the website with accurate information which is ranked at the top. It indentifies trustworthy websites better than popular search engines. Srikantaiah K C et al.,[18] propose a new technique called PCF engine to find the accuracy of the facts provided by the websites. It uses the probability based similarity function which performs the string matching between the true facts and the facts of web pages to find their probability of correctness.

## 3. Background

Xiaoxin Y et al., [1] propose a Truthfinder algorithm to find true facts with conflicting information from different infomation providers on the web. This approach is applied on certain domain such as, book authors and Movie run time. For the books domain, the Truthfinder uses author name as the facts which assigns the weights for first, middle and last name of the authors to determine the confidence of the facts and this is repeated for every fact to find trustworthiness of the website. It assigns the weight ratio of 2:1:3 for first, middle and last name respectively.

*Example:* True fact says, the author of some book is Graeme C. Simsion, where weight 2 is assigned for Graeme, 1 for C and 3 for Simsion and if the fact obtained from book seller website is Grame Simsio, where it does not contain middle name C and the charater 'n' is missing in the last name, it is only partially correct. Truthfinder assigns the half of the weight allocated for last name, i.e., 3/2 and full weght of 2 to first name and zero to middle name, therefore confidence of the fact is (2+1.5)/6 which is 58.33%.

PCF-Engine performs string matching between author names provided by the book sellers with author names of the corresponding book in the knowledgebase and hence, it searches for exactness of the fact and confidence of the above example is 93.75%. So, it is more accurate than Truthfinder.

## 4. Proposed Model

4.1 Problem Definition

Consider given a set of objects, a set of websites providing conflicting facts for an object and a set of true facts for a specific domain, the main goal is to rank the webpages providing the facts by finding the Probability of Correctness of the conflicting facts with respect to the true facts.

Assumption: The facts available in Knowledge base are 100% accurate and it is obtained from the trustworthy resource.

4.2 Mathematical Model

*Basic Definitions*

(i) *Probability of Correctness of Fact (PCF)*- is defined as the probability by which the fact is similar to that of the true fact or in other words by the factor that the fact has minimal deviation from the true fact.
(ii) *Implication between the facts* - is defined as the extent by which the facts has influenced other facts of the same object, i.e., the deviation between the PCFs of the facts from the threshold (maximum allowable deviation between the PCFs of the facts).

Trustworthiness of website is directly proportional to confidence of all the facts provided by that website and implication between the facts[1]. The basic notations used in the model are shown in Table 1.

Table 1: Basic Notaions

| Symbol | Description |
|---|---|
| $\varepsilon$ | : is the threshold, i.e., allowable deviation of PCFs between any two facts. |
| $p(f)$ | : is the probability of correctness of the fact about an object in some Domain. |
| $\Delta$ | : Difference between $p(f_1)$ and $p(f_2)$. |
| $Ob\{\}$ | : Set of objects in certain domain. |
| $TF\{\}$ | : Set of true facts indexed by objects. |
| $F'\{\}$ | : Set of facts provided by different websites indexed by objects. |
| $Web\{\}$ | : Set of websites URLS indexed by objects. |
| $s(f)$ | : Confidence of a fact $f$. |
| $s'(f)$ | : Adjusted confidence of a fact $f$. |
| $\sigma^*(f)$ | : Adjusted confidence score of a fact $f$. |
| $Inf(f_1, f_2)$ | : Influence between two facts $f_1$ and $f_2$. |

4.2.1 *Probability of Correctness of Facts (PCF)*
If $\exists f \in TF\{\}$, such that $f \leftarrow TF\{o\}$, provided by $Web\{o\}$, where, $\forall o \in Ob\{\}$ then,

$$p(F'_i\{o\}) \leftarrow SIM(f, F'_i\{o\}). \quad (1)$$

where, $1 \leq i \leq |F'\{o\}|$, $SIM(f, F')$ is defined as the factor by which $F'$ is true with respect to $f$ and it is based on the domain or context where it is used. If $F'$ is completely true with respect to $f$, then the probability of $F'$ is correct when

$f$ is considered as true fact is 1 i.e., $F'$ is 100% correct about an object, where $F'_i \{o\}$ is a $i^{th}$ fact for an object $o$ provided by some website as shown in Eq. (1). It implies that facts obtained from the website about an object is exactly similar to that of the true fact of an object and therefore the $SIM(f, F')$ is also used to find the initial trustworthiness of the website by applying this function between all the facts provided by the website and the corresponding objects true facts available in the knowledgebase.

### 4.2.2 Implication between Facts

Let $\Delta$ represents the difference between the probability of two facts $f_1$ and $f_2$, i.e., $\Delta = p(f_1) - p(f_2)$ and based on the value of $\Delta$, there are three cases.

*Case 1:* If $(0 < \Delta < \varepsilon)$ or $(\Delta > 0$ and $\Delta > \varepsilon)$ then, $f_1$ has low impact on $f_2$ by $|\varepsilon - \Delta|$.
Example: if $p(f_1)=0.7$ and $p(f_2)=0.2$ then $\Delta=0.5$ which is greater than $\varepsilon$, i.e., 0.4, this implies $f_1$ is 70% correct and $f_2$ is 20% correct, the difference is 50% which is greater than 40%(threshold) which is preferably allowed deviation between any two facts, therefore $f_1$ is having low impact on $f_2$ by 10%(50-40)%.

*Case 2:* If $(\Delta > 0)$ and $(\Delta = \varepsilon)$; then, $f_1$ has impact of $\varepsilon$ on $f_2$. The difference between the probabilities of correctness of the facts is equivalent to the value of threshold and hence $f_1$ has the impact $\varepsilon$ on $f_2$.

*Case 3:* If $(\Delta < 0)$ then, $f_1$ has high impact on $f_2$.
Example: if $p(f_1)=0.2$ and $p(f_2)=0.7$, then, $\Delta = -0.5$ which is negative and less than $\varepsilon$, i.e., 0.4, this implies $f_1$ is 20% correct and $f_2$ is 70% correct, the difference is -50% which is less than 40%(preferably allowed deviation between any two facts), therefore $f_1$ is having high impact on $f_2$ by 90% (40-(-50))%. In otherwords, by adding 50% to $f_1$ gives $f_2$ correctness, therefore $f_2$ is having low impact on $f_1$. Hence impact or influence between any two facts $f_1$ and $f_2$ on the same object can be defined as,

$$Inf(f_1, f_2) = \begin{cases} c(f_2) & \text{for case 1 and 3} \\ \varepsilon & \text{for case 2.} \end{cases} \quad (2)$$

where, $c(f_2) = |\varepsilon - \Delta| * s(f_2)$. Adjusted confidence of a fact is defined as,

$$s'(f_1) = s(f_1) + \sum_{o(f_1)=o(f_2)} Inf(f_1, f_2). \quad (3)$$

where, $s(f)$ is a confidence of a fact $f$ defined in [1]

$$s(f) = 1 - \pi(1 - t(w)). \quad (4)$$

$$s'(f) = min \begin{cases} c'(f) : c'(f) > 1, \\ 1 \leq \alpha \leq \infty \end{cases} * 10^{-\alpha} \quad (5)$$

where, $c'(f)=s'(f)*10^{-\alpha}$ and adjusted confidence score is defined in [1]

$$\sigma^*(f) = -ln(s'(f)). \quad (6)$$

In Eq. (5) dumping factor i.e., $10^{-\alpha}$ is multiplied to the adjusted confidence $s'(f)$ to get the probability value less than or equal to 1.

### 4.2.3 SIM(TF,F') for Books Domain

Let $Ob = \{ob_1, ob_2, ob_3, \ldots, ob_n\}$, $TF = \{TF_{11}, TF_{22}, TF_{33}, \ldots, TF_{nn}\}$ and $F=\{F'_{11}, F'_{22}, F'_{33}, \ldots, F'_{nn}\}$. where,

$$F'_{ij} = \{y_{ik} : 1 \leq k \leq n_b\}. \quad (7)$$

where, $n_b$ is the number of authors in the $i^{th}$ fact about the $j^{th}$ object and $i=j$. $F'_{ij}$ is again the set of authors for $j^{th}$ object (book). For example: $F'_{22}= \{y_{21}, y_{22}, y_{23}, \ldots, y_{2nb}\}$ is the set of authors of the second book(second object). The true fact can also contain only one author or a set of authors for a book as defined according to Eq. (8),

$$TF_{ij} = \{x_{ik} : 1 \leq k \leq n_a\}. \quad (8)$$

where, $n_a$ is the number of authors in the $i^{th}$ true fact about the $j^{th}$ object and $i=j$ and $x_{ik} = TF_i - X$, $X = x_l : 1 \leq l \leq n_a$ and $l! = k$. Therefore, the similarity function between $i^{th}$ true fact and corresponding $i^{th}$ fact provided for any object $o \in Ob$ is defined according to the Eq. (9),

$$SIM(TF_i, F'_{io}) = \sum_{j=1}^{|F'_{io}|} \sum_{k=1}^{|TF_{io}|} \frac{L(F'_{ioj})}{L(TF_{iok})}. \quad (9)$$

if and only if $F'_{ioj} \subseteq TF_{iok}$. Repeat the process for all object $o \in Ob$, therefore

$$SIM(TF, F') = \sum_{i,o=1}^{|F'|} \sum_{j=1}^{|F'_{io}|} \sum_{k=1}^{|TF_{io}|} \frac{L(F'_{ioj})}{L(TF_{iok})}. \quad (10)$$

$L(f)$: Gives the number of characters found in the author name F' in Eq. (9) and Eq. (10). Here, the probability of correctness is calculated depending on the number of characters matched in the first, last and middle of the author name of obtained facts from different website to the first, last and middle name of the author names taken in that order of the true fact about the object(book) available in the knowledge base.

*Example*: If $TF=\{\{Cay\ S\ Horstmenn, Gary\ Cornell\}, TF_{22}, \ldots, TF_{nn}\}$ is a true fact about the book Core java Volume 1 with ISBN 8131701621 and $TF_{11}=\{Cay\ S\ Horstmenn,\ Gary\ Cornell\}$, $F'=\{\{Cay\ S\ Horstmenn,\ Gary", "Horstmenn",\ "Corne\}, F'_{22}, \ldots, F'_{nn}\}$ where, $F'_{11}=\{Cay\ S\ Horstmenn,\ Gary,\ Horstmenn,\ Corne\}$.

Consider first author $F'_{111}$ i.e., Cay S Horstmenn provided by $w_1$ which is same as the author name provided in true fact Cay S Horstmenn therefore $p(F'_{111})=1$. Similarly, consider the second author $F'_{112}$ i.e., Gary which is a subpart of true fact Gary Cornell therefore, $p(F'_{112})=$ LEN(Gary)/LEN(Gary Cornell) which is 0.33 similarly $p(F'_{113})=$ 0.6 and $p(F'_{114})=0.416$. Therefore, initial trustworthiness of $w_1$, $t(w_1)=$ (1+0.33)/2 i.e., 0.625 on ISBN 8131701621; if w1 provides $F'_{111}$ and $F'_{112}$. Similarly, initial trustworthiness of $w_2$ on ISBN 8131701621 is $t(w_2)=1.016/2$, i.e., 0.508 where, 2 indicates number of facts provided by websites on the object(ISBN 8131701621) if it provides $F'_{113}$ and $F'_{114}$. This process is repeated for every object(every book) provided by the corresponding websites to get their respective initial trustworthiness and they are ranked accordingly.

Initially, it is assumed that none of the websites are trustworthy, therefore initial trustworthiness of all websites are assigned to zero. Therefore, the trustworthiness of website $t(w)$ is redefined as,

$$t(w) = \begin{cases} SIM(TF, F'), & if\ t(w) = 0 \\ \sum_{f \in F(w)} \frac{s(f)}{|F(w)|} & : otherwise \end{cases}[1].$$

(11)

where, $F'$ is a set of facts provided by website $w$ and $F' \subseteq F$. If trustworthiness is zero, then the website is added with the new data to the database whose trustworthiness is calculated using $SIM(TF, F')$, $TF$ indicates the true facts that is present in knowledge base which gives the probability of correctness of the fact $F'$ with the $TF$ of some object $o$ belonging to $Ob$, this is repeated for every object. Otherwise, trustworthiness is calculated by taking the average of confidence of all the facts provided by the website w in Eq. (11).

The proposed *Probability of Correctness of Fact* (PCF) engine ranks the page depending on the accuracy of the facts provided by the websites. The facts which are assumed as true about any object are stored in knowledge base. For example the true facts about the different books are taken from the respective coversheets of the books. Following properties are some of the facts taken for the book, ISBN: Uniquely identifies the fact, Author Names: Authors for the corresponding book, Publisher: publisher for the book, Price: cost of the book. Once knowledge base is constructed, the dataset containing conflicting facts for the various objects are populated using the website *www.abebooks.com*.

The $\varepsilon$ is set to 0.4 which indicates that 40% deviation in PCF between the facts are allowed. The algorithm behavior can be rendered by changing the value of threshold. The algorithm includes three important stages: *(i)* calculation of trustworthiness of all the websites, *(ii)* calculation of confidence of all the facts available in database and *(iii)* finding the influence between the facts. Since the algorithm operates on real dataset it is scheduled to run on every day to update the contents of the database. The Initial trustworthiness is calculated depending on the PCF of all the facts provided by the website where the PCF for every fact is determined by using Probability based similarity function (*SIM*). The facts provided by the different websites may be similar to the true fact and hence the PCF for those facts is 1. Which indicates the fact is 100% true and this is calculated on a fly in a single iteration. If PCF of all the facts provided by website is 1 and the deviations in the implications between the facts are low, then the trustworthiness closely approaches to 1 and hence the PCF engine always probes for exactness of the facts about an object.

The algorithm calculates trustworthiness for every websites by finding the PCF for the facts and it also computes the confidence of the facts by taking the trustworthiness of the corresponding websites providing the facts and hence trustworthiness and confidence are totally depending on each other. The algorithm stops after computing the trustworthiness of all the websites and confidence for all the facts found in the database. It recomputes the trustworthiness and confidence values when it is scheduled for next execution by considering the new facts arrived after the previous Execution. The algorithm is presented in Table 2.

## 5. Complexity Analysis

5.1 Time Complexity

The PCF-similarity operation is considered as basic operation for the above algorithm which is performed once for every website. Therefore, time complexity is O(*N*) where, $N = |W|$ and the time complexity for PCF-similarity function is O(*LMK*) where, $L = |F'|$, $M = |F'_{io}|$ and $K = |TF|$ for every object $o$. Therefore, $f(L, M, K, N, O_n) =$

$$\begin{cases} O(NLKM), & if\ t(w) = 0, 1 \leq w \leq N \\ O(O_n KM), & for\ all\ object\ o, 1 \leq o \leq O_n \end{cases}$$

(12)

Table 2: Algorithm for PCF-Engine

```
Input
  TF{}  : Set of true facts indexed by objects.
  F'{}  : Set of facts provided by different websites indexed by objects
  Ob{}  : Set of objects.
  Web{}: Set of Websites URL'S providing the facts.
  ε     : Equal to 0.4, allowable deviation between any two facts on the
          same domain.
Output
  Trustworthiness of the websites and confidence of the facts.
Process
begin
  for each w ∈ Web ; Compute Trustworthiness for every Website
    do
      if( t(w)==0 ) then
        t(w)= SIM(TF, F') ; where F' is the set of facts provided
                          ; by website w
      else
        t(w)= ∑_{f∈F(w)} s(f)/|F(w)|
      end if
  end for
  for each f ∈ F ; Compute the Confidence of the facts
    do
      s(f)=1-π(1-t(w)) ; for every website w providing a fact f
                       ; where, w ∈ Web{}
      σ(f)=-ln(1-s(f)) ; confidence score of a fact f
  end for
  for each f ∈ F ; Compute the Implication between the facts
    do
      for each f' ε F and f' ≠ f
        do
          Δ=p(f) - p(f')
          if Δ=ε then
            inf(f, f')+= ε
          else
            inf(f, f')+= | ε − Δ | * s(f')
          end if
      end for
      s(f) ← get the confidence of f from database
      s'(f)=s(f)+∑_{o(f')=o(f)} inf(f, f')
  end for
end
```

Since, *SIM(TF, F )* is executed for every website during the calculation of initial trustworthiness the time complexity is O(*NLKM*) and this happens only if initial trustworthiness is zero. But implication between the facts are calculated every time when algorithm is scheduled for execution and moreover number of objects facts are more than number of websites, the time complexity is, *f(L, M, K, N, O_n)* = max{O(*NLKM*), O(*O_nKM*)} and the number of times the *SIM(TF, F )* is executed is given by,

$$f(L, M, K, N, O_n) = O(O_n MK) \qquad (13)$$

where, $NLM \leq O_nM$. If $t_{op}$ is the time required to perform *SIM(TF, F )* then the time complexity is,

$$PCF\_TIME = O(O_n KM) * t_{op} \qquad (14)$$

$t_{op}$ includes the time required to access the contents from database. If $t_{op}O_n$ is the time required to access $O_n$ number of facts about the object and if $t_{op}w_n$ is the time required to access the facts provided by $w_n$ number of websites then total time required is,

$$t_{op} = t_{op}O_n + t_{op}w_n \qquad (15)$$

5.2 Space Complexity

The space complexity includes the total space required to store knowledge base and database containing the facts provided by the websites. This is defined as,

$$s(O_n, O_m, W_n) = O(O_n) + O(w_n O_m). \qquad (16)$$

O($O_n$) is the space required to store true facts about $O_n$ number of objects in knowledge base. O($w_n O_m$) is the space required to store conflicting facts provided by $w_n$, if each website provides facts for $O_m$ number of objects.

## 6. System Architecture

The entire functionality is depicted in Fig. 1, which consists of the following activities,
1. User inputs the keyword for the search.
2. The web server communicates with the PCF engine to

perform the similarity analysis between true facts and conflicting facts obtained from different information providers.
3. A PCF engine extracts true facts from the knowledgebase as interestingness measure.
4. PCF Engine also extracts conflicting facts available in database to perform similarity analysis with true facts.
5. The search engine in web server uses the results of PCF-similarity function to rank the websites.
6. The result of the search engine is returned to user(Browser).

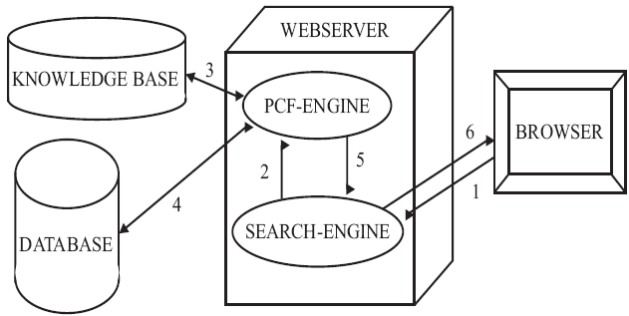

Fig. 1 System Architecture

Knowledge Base : The knowledgebase contains true facts about the objects which are obtained from the known trustworthy source. The true facts are used as interestingness measure to apply PCF-similarity function between them and facts available in database. The true facts about the objects are highly domain specific. *Example:* true facts about books are different from true facts about vehicle.

DataBase : The database contains conflicting facts provided by different information providers in the web. This database includes URL of the information providers and their corresponding trustworthiness. Similarly, facts provided by these information providers are stored with their corresponding confidence and confidence score. Note that both trustworthiness and confidence values are probability values which are less than or equal to 1.

PCF − Engine : Probability of correctness of the facts-Engine is the vital component of the system which performs relevance or similarity analysis between true and conflicting facts.

The entire workflow of PCF-Engine is depicted in the Fig. 2, where, true facts and conflicting facts from the knowledge base and database form an input to the PCF-Engine. The PCF-Engine assumes that every website is not trustworthy and hence the facts provided by the websites are not true, therefore it initializes the trustworthiness, trustworthiness score, confidence and confidence score to zero. For every website the PCF-Engine accumulates the facts provided by the website to apply PCF-similarity function to perform the relevance analysis between true facts and accumulated facts and it updates the corresponding trustworthiness of the website. The PCF-similarity function is applied on the facts of the website only during calculation of initial trustworthiness else it is calculated as the average of confidence of the facts. The next activity performed by PCF-Engine is calculation of confidence of the facts. The PCF-Engine accumulates trustworthiness of the websites providing the fact to calculate confidence of the fact.

The PCF-Engine uses the PCF-similarity function to find the implication between facts. The implication is determined using the deviation found between PCF of the facts. The confidence of fact is adjusted by finding implication of the fact on all facts available for the corresponding object. The facts can be provided by any website and hence the implication affects the trustworthiness of website when PCF-Engine is scheduled to run for the next time. Therefore, trustworthiness of the websites not only depends on confidence of the facts provided but also depends on implication between the facts because while calculating an implication the facts are categorized according to the objects irrespective of the websites providing them. When calculating the trustworthiness for the website the facts are grouped according to the website.

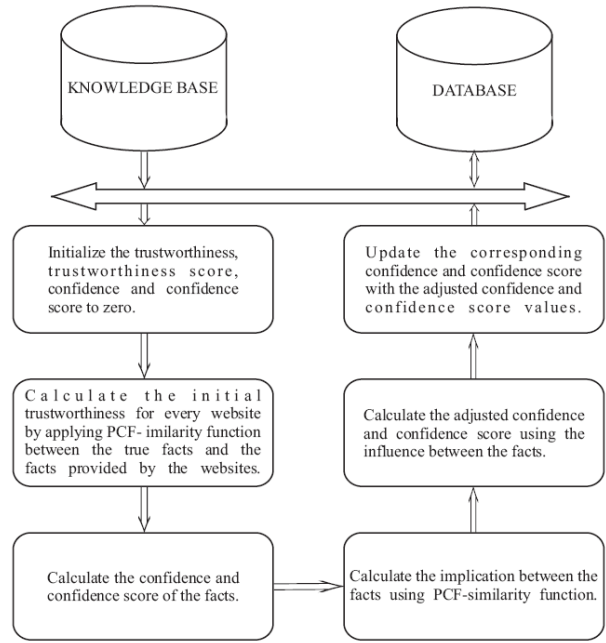

Fig. 2 Workflow Inside PCF-Engine

Finally, confidence and confidence score of the facts are updated by using the adjusted confidence and confidence score values and it is updated accordingly in database. Note that knowledgebase reads only database for PCF-Engine. The Knowledge base is updated by administrator when new true facts are to be added.

Web Server : The web server is responsible for querying the database to rank the websites according to trustworthiness of the websites. When it receives keyword for a search it communicates with the PCF-Engine to update trustworthiness, trustworthiness score, confidence and confidence score by using PCF-similarity function. The search engine logic and PCF-engine logic are executed by the web server as shown in Fig. 1.

## 7. Experimental Results

The data set consists of facts for the books domain, where domain in this context corresponds to values of certain attributes of the book such as ISBN, Author Names, Publisher, Price, URL of Book seller website and quantity (availability). The data set consists of the above specified information for 50 websites with 100 facts. The initial trustworthiness of websites, confidence and confidence scores of all the facts are initialized to zero. The Author Name of the book is considered as the important fact for Probability based similarity function (*SIM*) to perform the relevance analysis. The result of the Probability based similarity function (*SIM*) for all the facts provided by a website is used to calculate the trustworthiness of the website and this is performed on all the websites to rank them accordingly.

Table 3: Trustworthiness Values of Different Websites.

| URL | Trustworthiness | Webid |
|---|---|---|
| http://www.abebooks.com/william bookstore-new-delhi/6815050/sf | 0.25 | 1 |
| http://www.abebooks.com/123-books-fremont-ca-u.s.a/ 53834279/sf | 0.210526315789474 | 2 |
| http://www.abebooks.com/balsingh-new-delhi/ 51068449/sf | 0.337938596491228 | 3 |
| http://www.abebooks.com/textbook-store-philade-lphia-pa-u.s.a/53227415/sf | 0.130756578947368 | 4 |
| http://www.abebooks.com/ aggarwal overseas-new-delhi/52327687/sf | 1 | 5 |
| http://www.abebooks.com/ a-c-m-books beaumaris/50192413/sf | 0.58333333 | 6 |
| http://www.abebooks.com/ a-i-pedersenmacclesfield/660873/sf | 1 | 7 |

The *PCF-Engine* is developed using ASP.NET with C# as the underlying language. The environment used for development inludes Visual Studio 2005 (IDE), Windows XP(OS) and the MySQL 4.0 for database. The initial implementation is done with the " set 0.4. The calculated trustworthiness values of 7 websites out of 50 websites are given in the Table 3. From the Table it is observed that the trustworthiness values of the booksellers a-i-pedersen-macclesfield (WebId 7) and aggarwal-overseas-new-delhi (WebId 5) are 1, which indicates that the facts provided by them are exactly similar to the true facts available in the knowledgebase.

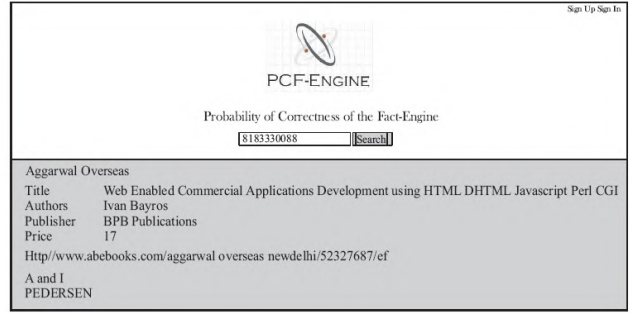

Fig. 3 Snapshot of the PCF-Engine

The trustworthiness values are calculated based on the confidence of all the facts provided by the corresponding websites on some object. As shown in Fig. 3, the search is made for the ISBN 8183330088 of the book titled with *Web Enabled Commercial Applications Development using HTMLDHTML Javascript Perl CGI*, the webIds providing the facts about this book are 5, 7, 3, 10, 8, 9 . . . etc., of which the most trustworthy websites with value 1 are 5 and 7 and they occupy the first two positions in the searched result.

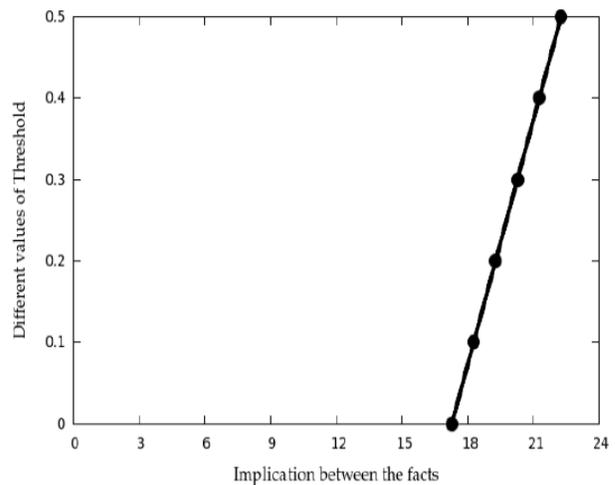

Fig. 4 Variation of Implication for Different Values of Threshold.

The behavior of the PCF-Engine is analyzed by observing the variations of the implication of the facts with respect to the change of " from 0.5 to 0. The zero value of " indicates that the influence between the facts is difference between the PCFs of the facts else it is the factor by which PCFs of the facts deviates from threshold (") according to the Eq. (2). The behavior of Eq. (2) is plotted in graph as shown in the Fig. 4, which indicates that influence between the facts decreases linearly with decrease in threshold value.

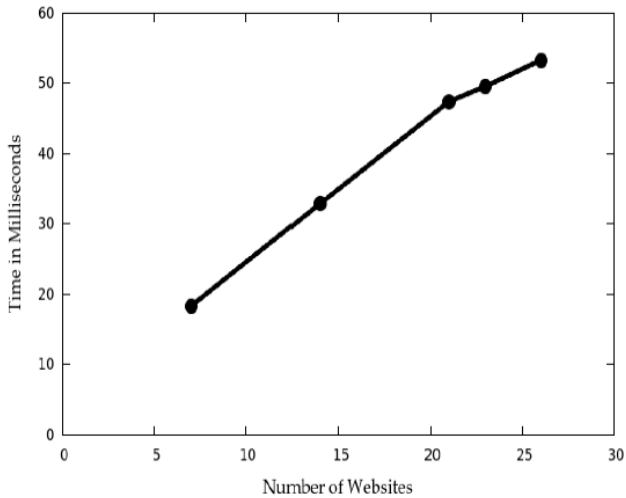

Fig. 5 Variation of PCF-Engine Time with Number of Websites.

The search time heavily depends on the time taken by the PCF-similarity function for finding the trustworthiness values of the websites. The time required to perform PCF-similarity function linearly increases with increase in the number of websites in the database as shown in the Fig. 5, Where the X-axis indicates Number of websites and Y-axis indicates PCF-time.

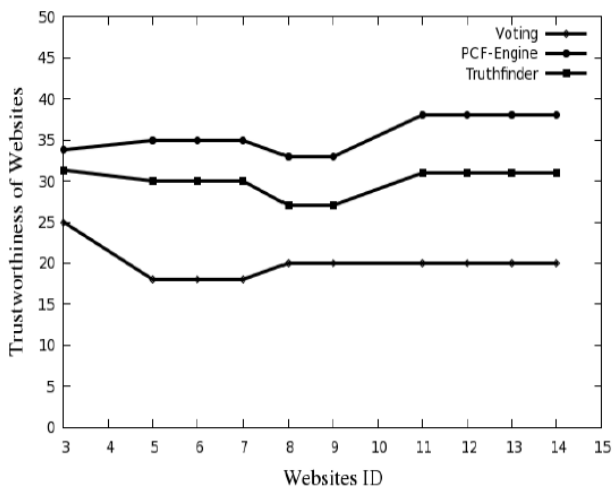

Fig. 6 Comparison Between the Trustworthiness Values of Truthfinder, Voting and PCF-Engine.

The graph is plotted for the websites providing the facts for the book *WebEnabled Commercial Application devlopment using HTML, DHTML, JavaScript, Perl, CGI" by Ivan Bayross*. As it is observed from the Fig. 6, the trustworthiness of websites fall in the range 18%-25% for Voting, 26%-33% for Truthfinder and 33%-38% for PCF-Engine and deviation between PCF Engine and Truthfinder with trustworthiness calculations is 0.058 and hence PCF-Engine is 5.8% more accurate than the Truthfinder. The probability values are normalized to two digits numbers in Y-axis. Since the Voting uses the facts count without considering the truthness of the facts provided by the websites, its accuracy is low compared to PCF-Engine and Truthfinder algorithms.

## 4. Conclusions

In this paper a new approach called PCF-Engine which uses Probability based similarity function (*SIM*) is proposed for resolving the conflicts between the facts provided by different information providers in the web. The Probability based similarity function (*SIM*) finds the implication between the facts. If the websites provides the fact which is exactly similar to that of true fact in the knowledge base the PCF-Engine computes its trustworthiness value as 1 on a fly in a single iteration. The work can be extended by dynamically fetching the true facts to the knowledge base and removing the domain specific dependency of true facts.

**Srikantaiah K C** is an Associate Professor in the Department of Computer Science and Engineering at S J B Institute of Technology, Bangalore, India. He obtained his B.E and M.E degrees in Computer Science and Engineering from Bangalore University, Bangalore. He is presently pursuing his Ph.D programme in the area of Web mining in Bangalore University. His research interest is in the area of Data mining, Web mining and Semantic Web.

**Srikanth P L** is a Post Graduate Student of Computer Science and Engineering, University Visvesvaraya College of Engineering, Bangalore University, Bangalore. His research interest is in the area of Web Technology, Semantic Web and Cloud Computing.

**Tejaswi V** is a Student of Computer Science and Engineering from Rastriya Vidyalaya College of Engineering, Bangalore. Her research interest is in the area of Wireness Sensor Networks.

**Shyla K** is an Asst. Professor in the Department of Electronics and Communication Engineering at Vivekananda Institute of Technology, Bangalore, India. She obtained her B.E and M.E degrees in Electronics and Communication Engineering from Bangalore University, Bangalore. she is presently pursuing her Ph.D programme in the area of Wireless Sensor Networks in Bangalore University. Her research interest is in the area of Sensor Networks, Adhoc Networks and Image Processing.

**K R Venugopal** is currently the Principal, University Visvesvaraya College of Engineering, Bangalore University, Bangalore. He obtained his Bachelor of Engineering from University Visvesvaraya College of Engineering. He received his Masters degree in Computer Science and Automation from Indian Institute of Science Bangalore. He was awarded Ph.D. in Economics from Bangalore University and Ph.D. in Computer Science from Indian Institute of Technology, Madras. He has a distinguished academic career and has degrees in Electronics, Economics, Law, Business Finance, Public Relations, Communications, Industrial Relations, Computer Science and Journalism. He has authored 31 books on Computer Science and Economics, which include Petrodollar and the World Economy, C Aptitude, Mastering C, Microprocessor Programming, Mastering C++ and Digital Circuits and Systems etc.. During his three decades of service at UVCE he has over 250 research papers to his credit. His research interests include Computer Networks, Wireless Sensor Networks, Parallel and Distributed Systems, Digital Signal Processing and Data Mining.

**L M Patnaik** is an Honorary Professor in Indian Institute of Science, Bangalore. He was a Vice Chancellor, Defense Institute of Advanced Technology, Pune, India. During the past 35 years of his service at the Institute he has over 700 research publications in refereed International Journals and Conference Proceedings. He is a Fellow of all the four leading Science and Engineering Academies in India; Fellow of the IEEE and the Academy of Science for the Developing World. He has received twenty national and international awards; notable among them is the IEEE Technical Achievement Award for his significant contributions to High Performance Computing and Soft Computing. His areas of research interest have been Parallel and Distributed Computing, Mobile Computing, CAD for VLSI circuits, Soft Computing and Computational Neuroscience.